\documentstyle[12pt,psfig]{article}
\renewcommand{\baselinestretch}{1.2}
\newcommand{\beq}{\begin{equation}}
\newcommand{\eeq}{\end{equation}}
\newcommand{\beqa}{\begin{eqnarray}}
\newcommand{\eeqa}{\end{eqnarray}}
\newcommand{\beqar}{\begin{eqnarray*}}
\newcommand{\eeqar}{\end{eqnarray*}}
\newcommand{\eg}{{\it e.g.,}\ }
\newcommand{\ie}{{\it i.e.,}\ }
\newcommand{\labell}[1]{\label{#1}}
\newcommand{\labels}[1]{\label{#1}}
\newcommand{\reef}[1]{(\ref{#1})}
\def\IR{{\hbox{{\rm I}\kern-.2em\hbox{\rm R}}}}
\begin{document}

\begin{titlepage}

\begin{flushright}
DTP 99-37\\
EHU-FT/9909\\
hep-th/9906040
\end{flushright}
\vfil\vfil

\begin{center}

{\Large \bf AdS/CFT Duals of Topological Black Holes and the Entropy of 
Zero--Energy States}

\vspace{25pt}

Roberto Emparan\footnote{roberto.emparan@durham.ac.uk}
\vfil

{\sl Department of Mathematical Sciences}\\
{\sl University of Durham, Durham DH1 3LE, UK}\\
and\\
{\sl Departamento de F{\'\i}sica Te\'orica}\\
{\sl Universidad del Pa{\'\i}s Vasco, Apdo. 644, E-48080, Bilbao, Spain}
\vfil

\end{center}

\vspace{5mm}

\begin{abstract}
\noindent 

The horizon of a static black hole in Anti-deSitter space can be spherical,
planar, or hyperbolic.  The microscopic dynamics of the first two classes of
black holes have been extensively discussed recently within the context of the
AdS/CFT correspondence.  We argue that hyperbolic black holes introduce new and
fruitful features in this respect, allowing for more detailed comparisons
between the weak and strong coupling regimes.  In particular, by focussing on
the stress tensor and entropy of some particular states, we identify unexpected
increases in the entropy of Super--Yang--Mills theory at strong coupling that
are not accompanied by increases in the energy.  We describe a highly 
degenerate
state at zero temperature and zero energy density.  We also find that the
entanglement entropy across a Rindler horizon in exact AdS$_5$ is larger than
might have been expected from the dual SYM theory.  Besides, we show that
hyperbolic black holes can be described as thermal Rindler states of the dual
conformal field theory in flat space.

\end{abstract}

\vfil\vfil\vfil
\begin{flushleft}
May 1999
\end{flushleft}
\end{titlepage}
\setcounter{footnote}{0}

\newpage
\renewcommand{\baselinestretch}{1.2}  

\section{Introduction}

The correspondence between string theory in Anti-deSitter (AdS) space and
conformal field theory (CFT) \cite{juan,gkp,wit1} provides a powerful basis for
the study of the microscopic statistical mechanics of black holes.  In this
framework, a black hole in AdS is described as a thermal state of the dual
conformal field theory\footnote{Small spherical black holes in AdS, however, 
are
unstable, and their entropy is not an extensive quantity.}.  The latter is
defined on a background geometry that is conformally related to the geometry at
the boundary of the AdS space.  If we want to work in a regime where the
supergravity approximation to string theory is reliable, then the dual CFT has
to be strongly coupled.  The aim of this paper is to develop the duality for a
class of black holes peculiar to AdS space, that will exhibit new and 
remarkable
features.

It is known that the presence of a negative cosmological constant allows for
more varied types of horizon geometries than in asymptotically flat situations.
In AdS the horizon of a black hole can have positive, zero, or negative
curvature.  These are spherical, planar or hyperbolic black holes, 
respectively.
In four dimensions it is possible to construct horizons of arbitrary topology 
by
modding out discrete isometry groups.  This is the origin of the name
``topological black holes."  We keep this name, even if it will be somewhat of 
a
misnomer since we will not be considering identifications under discrete
isometries.  Nevertheless, that is something that could be implemented in a
straightforward manner.

The microscopic study of planar black holes within string theory can be traced
back to the discussion in \cite{guklpe} of the statistical mechanics of black
D3-branes.  As this system is understood now, the planar black hole in AdS$_5$
is dual to a thermal state of ${\cal N}{=}4$ supersymmetric Yang--Mills (SYM)
theory in four dimensional Minkowski space, with gauge group $SU(N)$, in the
large $N$ limit and at a large value of the `tHooft coupling $g^2_{YM}N$.  We
have very limited knowledge of gauge theory in such a strong coupling regime,
but the results that follow from calculations using AdS supergravity appear to
be remarkably close to what we are able to compute using free field theory.  
The
AdS$_5$/SYM pair is the most studied case, but for other dimensions we know 
that
the temperature dependence of the dual field theories is determined by 
conformal
invariance, and this behavior is indeed reproduced by planar black holes
\cite{klts,wit2}.

Spherical black holes, on the other hand, present a different qualitative
feature, namely, a phase transition at finite temperature \cite{hp}.  As
observed in \cite{wit1,wit2} this phase transition fits in nicely with our
expectations of a confining phase at low temperatures for large $N$ theory on a
spatial sphere.  This is remarkable.  Confinement, however, is a phenomenon 
well
beyond the reach of perturbative field theory.  In the present paper, instead,
we will be more interested in situations where we can have some hope of 
connecting the weak and strong coupling regimes.

Hyperbolic black holes in the AdS/CFT context have received comparatively
little attention.  It was observed from their thermodynamics that the dual
field theories, defined on a spatial hyperboloid, should have no phase
transitions as a function of temperature \cite{re,birm}.  At any non--zero
temperature the theory is in a deconfined phase, and would appear to be free
from drastic changes of degrees of freedom as the coupling is increased. On
the other hand, the presence of the length scale coming from the curvature of
the hyperbolic space introduces a structure richer than in the case of flat
space.  There are two additional features of interest.  One of them is the fact
that the ground state is in general different from the solution that is locally
isometric to AdS.  In fact, the latter is a solution at finite temperature,
with non vanishing entropy, whose origin is due to the presence of a
non--degenerate (bifurcate) acceleration horizon in AdS.  A second aspect of
interest is that the boundary geometry is conformal to Rindler space.  It
follows that hyperbolic black holes admit a dual description as thermal Rindler
states of the CFT in flat space.

Perhaps the most startling consequence of our study will be that in the strong
coupling regime we are able to identify larger entropies than would be expected
from the CFT side.  The first example of this is the ground state in the
infinite coupling regime, which is shown to possess a large degeneracy, even if
it is a zero--temperature, zero--energy density state.  The next example we
describe is a supergravity state that is locally isometric to AdS$_5$, with an
entropy that turns out to be larger than expected from the calculation at weak
coupling.  Moreover, the increase in the entropy is not accompanied by an
increase in the energy of the state. We therefore find a common thread in these 
results, which would appear to point to the possibility that SYM theory
requires the presence of states that can give rise to an entropy, but do not
contribute to the local energy density.  Curiously, states with precisely these
properties have been postulated from a different analysis of the AdS/CFT
correspondence \cite{pst}, where the issue of causality in scattering processes
was studied.  Although it is probably to soon to discard other alternatives,
it would be really exciting if the two phenomena were related.

The layout of the paper is as follows:  Section \ref{sec:topobhs} introduces 
the
black holes under consideration, and their quasilocal stress-energy tensor and
entropy are presented.  Part of these results had been obtained in
\cite{birm,ejm}.  In section \ref{sec:review} we provide a review of the dual
CFT description of planar and spherical black holes with the focus on the
aspects that will change when we look at hyperbolic black holes.  The
supergravity and field theoretical descriptions of the latter are the subject 
of
detailed comparison in section \ref{sec:hyperduals}.  Section \ref{sec:rindler}
develops the description of hyperbolic black holes as Rindler states of the 
dual
CFT in flat spacetime.  In section \ref{sec:correc} we address the issue of
finite coupling corrections.  Finally, we discuss in section 
\ref{sec:precursor}
the possible identification of exotic states from this analysis.

\section{Topological black holes}\labels{sec:topobhs}

Our subject in this paper will be the following black hole solutions in 
AdS$_{n+1}$:
\beq
ds^2=-V_k(r)dt^2 +{dr^2\over V_k(r)}+{r^2\over l^2} 
d\Sigma^2_{k,n-1}\ ,
\labell{bhmetric}
\eeq
with
\beq
V_k(r) =k-{\mu\over r^{n-2}}+{r^2\over l^2}\ ,
\labell{vk}
\eeq
where the $(n{-}1)$ dimensional metric $d\Sigma^2_{k,n-1}$ is
\begin{equation}
d\Sigma^2_{k,n-1} =\left\{ \begin{array}{ll}
\vphantom{\sum_{i=1}^{n-1}}
l^2d\Omega^2_{n-1}& {\rm for}\; k = +1\\
\sum_{i=1}^{n-1} dx_i^2&{\rm for}\; k = 0 \\
\vphantom{\sum_{i=1}^{n-1}}
l^2dH^2_{n-1} &{\rm for}\; k = -1\ ,
\end{array} \right.
\labell{little}
\end{equation}
where $d\Omega^2_{n-1}$ is the unit metric on $S^{n-1}$. By $dH^2_{n-1}$
we mean the ``unit metric" on the $(n{-}1)$--dimensional hyperbolic space 
$H^{n-1}$.

The solutions for $k=+1$ are sometimes called ``Schwarzschild-AdS'' solutions:
They reduce to the standard Schwarzschild solution when the cosmological
constant vanishes, $l\rightarrow\infty$, and to AdS in global coordinates when
$\mu=0$.  Moreover, their topology is $\IR^2\times S^{n-1}$, and the horizon
is the sphere $S^{n-1}$, like that of the Schwarzschild solution.  The case
$k=0$ makes appearance when considering the near--horizon limit of
(non--dilatonic) $p$-branes.  Their horizon has the geometry of $\IR^{n-1}$,
which can be periodically identified to give horizons of toroidal topology,
although we will not consider such possibilities.  Both the $k=+1$ and the
$k=0$ cases have been extensively studied recently in the context of the
AdS/CFT correspondence.

By contrast, the class of hyperbolic solutions $k=-1$ have received
comparatively less attention.  They have been studied mostly in four
dimensions, where, together with the other two classes, they can be used to
construct black holes with horizons of arbitrary topology:  if the hyperbolic
space $H^2$ is identified under appropriate discrete subgroups of the isometry
group, then all the closed Riemann surfaces of genus higher than 1 can be
generated \cite{toporefs}.  A similar result holds for five-dimensional black
holes \cite{birm}, as follows from the fact that an arbitrary compact
three-manifold of constant curvature can be constructed as a quotient of a
universal covering space of positive, zero or negative curvature.  This is the
origin of their denomination as ``topological black holes."  Their appearance
in M-theory and in the context of the AdS/CFT correspondence was first
discussed, in four dimensions, in \cite{re}.  In higher dimensions they have
been studied first in \cite{birm}.

The temperature of these black holes is determined in the standard 
(Euclidean) 
manner as
\beq
\beta={4\pi l^2 r_+\over nr_+^2+ k(n-2)l^2}\ ,
\eeq
where $r_+$ is the horizon radius. This relation can be inverted to find
\beq
r_+={2\pi l^2\over n \beta}\left[1+\sqrt{1- k{n(n-2)\beta^2\over 4\pi^2 
l^2}}\; \right]
\labell{rplus}
\eeq
which allows us to take $\beta$ as the parameter that determines the 
solution\footnote{For $k=+1$ a second solution for $r_+$ exists, with a 
negative sign for the square root, which corresponds to small black holes 
in AdS. We will not discuss these.}. Notice 
that in the limit where $r_+\gg l$ the $k=\pm 1$ classes of solutions 
approach the planar black hole class $k=0$. This admits an interpretation in 
terms of an ``infinite volume" limit, in which the curvature radius of 
$S^{n-1}$ or $H^{n-1}$ is much larger than the thermal wavelength of the 
system \cite{wit2}. 

\begin{figure}[ht]
\hskip1cm
\psfig{figure=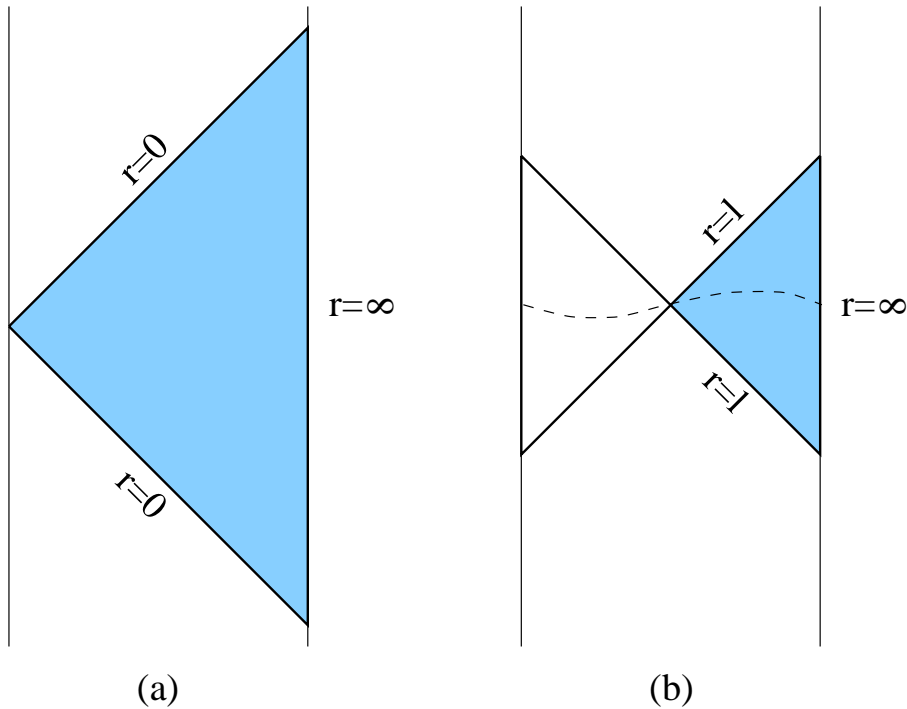}%
\hskip2cm
\caption{Regions of Anti-deSitter space covered by the parametrizations 
$k=+1,0,-1$ ($\mu=0$ in eq.~\reef{vk}). With $k=+1$ all of AdS (the entire 
strip) is covered. (a) Portion covered in Poincar\'e (or horospheric) 
coordinates, $k=0$. (b) Portion covered by the hyperbolic slicing $k=-1$. 
Shown 
as a dashed line is a Cauchy surface.}
\label{fig:patches}
\end{figure}

At this point it is worth recalling that the solutions for $\mu=0$ are all 
isometric to 
AdS$_{n+1}$, and therefore can be locally transformed into one 
another by a simple redefinition of coordinates\footnote{Other black-hole-type 
solutions constructed by performing identifications in AdS space have been 
considered in \cite{ccbhs}. However, they exhibit pathologies, not only within 
Einstein-AdS gravity \cite{ccbhs} but also within the context of the AdS/CFT 
correspondence  \cite{unp}.}. However, there are 
non--trivial 
differences between these parametrizations. The metric with $\mu=0$, $k=+1$ 
describes AdS in global coordinates, whereas $k=0$ describes the Poincar\'e 
(or horospheric) parametrization of AdS. The latter describes a wedge of AdS, 
since the coordinate system breaks down at $r=0$, see Fig.\ 
\ref{fig:patches}a. 
This coordinate singularity corresponds to a degenerate Killing horizon. This 
means that, in contrast to bifurcate Killing horizons, there is no temperature 
associated to it. Besides, its area vanishes. Then, a common feature of AdS in 
both its $k=+1$ and $k=0$ forms is the vanishing of entropy and temperature. 
They are to be thought of as the 
ground states of their respective classes of solutions.

The solution with $\mu=0$, $k=-1$, introduces a difference here.  While
isometric as well to AdS, it covers a smaller portion of the entire manifold,
as the coordinate patch breaks down at $r=r_+=l$, see Fig.\
\ref{fig:patches}b.  However, in contrast to the horizon in Poincar\'e
coordinates, the horizon in this case is analogous to a Rindler horizon.
There is an associated inverse temperature, $\beta=2\pi l$, and it has
non--vanishing area.  One should note that, among the $k=-1$ class of black
hole solutions, the one that is isometric to AdS is not properly a black hole.
It is completely non--singular, and in the absence of identifications it does 
not possess an event horizon. By contrast, the solutions with $\mu\neq 0$ 
possess a singularity at $r=0$.

For the $k=-1$ class of black holes, and in contrast to the 
$k=+1,0$ classes, the zero temperature solution is different from the one 
that is isometric to AdS. In fact, for $k=-1$ there is a range of negative 
values for 
$\mu$ such that the solutions still possess regular horizons. The minimum 
values of $\mu$ and $r_+$ that are compatible with cosmic censorship, for 
which 
the 
horizon is degenerate, are
\beq
\mu_{\rm ext} = -{2\over n-2}\left({n-2\over
n}\right)^{n/2} l^{n-2}\ ,\qquad r_{\rm ext}=\sqrt{n-2\over n} l\ ,
\eeq
and, in particular, 
\beq
\mu_{\rm ext}=-{l^2\over 4}\ ,\qquad r_{\rm ext}={l\over\sqrt{2}},\qquad {\rm 
for\;}
n=4\ .
\labell{ext4}
\eeq
For these values of the parameters, the black hole is extremal. The Penrose 
diagram for a hyperbolic black hole with negative $\mu$ is like that of a 
Reissner-Nordstr\"{o}m-AdS black hole. For positive $\mu$ it is instead like 
that of a Schwarzschild-AdS black hole \cite{toporefs}.

We now want to evaluate the thermodynamic functions for the solutions
\reef{bhmetric}.  In particular, if we have the quasilocal stress-energy 
tensor,
which is defined on the boundary of a region of spacetime \cite{bryo}, as a
function of the temperature then we can compute all other thermodynamic 
functions
such as the energy or entropy.  Recently, a prescription for computing the
quasilocal stress tensor of a solution in AdS space has been proposed which
appears to capture all of the information relevant to the dual field theory
\cite{pv}.  In this prescription, regularization does not proceed by the
traditional subtraction of similar divergences from a reference state to which
the solution is asymptotically matched.  Instead, in the regularization 
proposed
in \cite{pv} divergences are removed by subtraction of local counterterms at 
the
boundary, in a manner closely analogous to the subtraction of divergences in
field theory in curved spacetimes.  As such, it appears to be particularly
suitable for constructing the stress tensor of the dual CFT starting from a
supergravity solution (see also \cite{rob}).  This technique has been extended
and generalized in \cite{ejm} to all the dimensions of relevance for
string/M-theory.

The metric on the boundary of AdS, $h_{\mu\nu}$ ($\mu,\nu=0,\dots,n-1$), is 
conformally related to the background metric of the field theory 
$\gamma_{\mu\nu}$. The conformal factor diverges near the boundary. By the 
AdS/CFT correspondence, the quasilocal stress tensor 
for AdS supergravity $\tau_{\mu\nu}$ can be translated into the 
expectation value of the stress tensor of the dual field theory $\langle 
T_{\mu\nu}\rangle$, in the strong 
coupling regime, as \cite{rob}
\beq
\langle {T_\mu}^\nu [\gamma_{\kappa\lambda}]\rangle= 
\left({h\over\gamma}\right)^{1/2} {\tau_\mu}^\nu[h_{\kappa\lambda}]\ ,
\labell{tensors}
\eeq
where the limiting approach to the boundary is assumed.

For the cases at hand the calculation of ${\tau_\mu}^\nu$ is straightforward. 
The appropriate conformal factor is $(h/\gamma)^{1/2}=(r/l)^n$ (see eq.\ 
\reef{bdry} below) and we obtain\footnote{Even if the 
counterterms 
introduced in \cite{ejm} can in general cancel divergences only up to 
$n=6$, a result equivalent to \reef{stressbh} was argued in that paper to hold 
for generic $n$.}
\beqa
\langle {T_\mu}^\nu \rangle &=& {1\over 16\pi G l}\left( {2\epsilon^0_k\over 
n-1}+
{\mu\over l^{n-2}}\right){\rm diag}(1-n,1,\dots,1)\nonumber\\
&=&{1\over 16\pi G l}\left( {2\epsilon^0_k\over n-1}+
{r_+^n\over l^n}+k{r_+^{n-2}\over l^{n-2}} \right){\rm 
diag}(1-n,1,\dots,1)\ ,
\labell{stressbh}
\eeqa
where (see \cite{ejm})
\beq
\epsilon^0_k=(-k)^{n/2}{(n-1)!!^2\over n!}\qquad{\rm for\; even\; }n\ ,
\labell{eps}
\eeq
and $\epsilon^0_k=0$ for odd $n$. It is worth noting that the form of this 
stress tensor 
is that of a thermal gas of massless radiation.

For the particular case of AdS$_5$ (and any $k$) it will be useful to note 
that the result can be written in a compact form as
\beq
\langle {T_\mu}^\nu \rangle={\pi\over 16 G l}\left({r_+\over \beta}\right)^2  
{\rm 
diag}(-3,1,1,1)\ .
\labell{stressads5}
\eeq

The energy, given as a function of temperature through \reef{rplus}, can 
be read from \reef{stressbh} as
\beq
E_{BH}(\beta)={(n-1) V_{n-1}\over 16\pi G l}\left({r_+^n\over 
l^n}+k{r_+^{n-2}\over l^{n-2}} +{2\epsilon^0_k\over n-1}\right)\ ,
\labell{energy}\eeq
with $V_{n-1}$ the volume of $d\Sigma^2_{k,n-1}$, \ie the spatial volume 
of the field theory. With $E$ as a function of the temperature we can 
apply standard thermodynamic formulae to compute the entropy of the 
solution,
\beq
S_{BH}={V_{n-1}\over 4 G}\left({r_+\over l}\right)^{n-1}\ ,
\labell{entropy}
\eeq
which satisfies, as expected, the Bekenstein--Hawking area law.
We could equally well have computed the Euclidean action of the solutions 
and in this way obtain $\beta$ times the free energy $F$, from which the 
same values of $E$ and $S$ are recovered \cite{ejm}.

\section{CFT duals of spherical and planar black holes---a brief 
review}\labels{sec:review}

The AdS/CFT correspondence states that the full non--perturbative dynamics of
quantum gravity in a space that is asymptotic to AdS$_{n+1}$ can be formulated
in terms of a dual conformal field theory defined on the $n$-dimensional 
causal
boundary of the bulk spacetime.  As discussed in detail in \cite{ejm}, the 
issue
of what is the geometry of the boundary of a given solution is, to some 
extent,
open, since it depends on how the spacetime is sliced radially as one 
approaches
the boundary.  As an example, it was explicitly shown in \cite{ejm} how the
boundary of (Euclidean) AdS$_{n+1}$ can be chosen to be $S^n$, $\IR^n$, $H^n$,
$\IR\times S^{n-1}$, $\IR\times H^{n-1}$, and several other geometries.  We 
see
then that the duals of AdS quantum gravity are in general conformal field
theories defined on curved backgrounds with {\it fixed} geometry.

More specifically, in the coordinates chosen in \reef{bhmetric}, the metric at 
the boundary, as $r\rightarrow\infty$, is of the form
\beq
h_{\mu\nu}dx^\mu dx^\nu\rightarrow {r^2\over l^2}(-dt^2 + d\Sigma^2_{k,n-1})\ .
\labell{bdry}
\eeq
The background spacetime for the dual field theory, $\gamma_{\mu\nu}$, is 
conformally related to this one, 
and 
the conformal factor can be chosen to cancel the divergent factor $r^2/l^2$ in 
\reef{bdry}, $\gamma_{\mu\nu}=\lim_{r\rightarrow\infty}{l^2\over r^2} 
h_{\mu\nu}$. In this way, the 
$k=+1,0,-1$ black holes admit a dual description 
in terms of a CFT on, respectively, $\IR\times S^{n-1}$, $\IR^n$, $\IR\times 
H^{n-1}$, each of these otherwise known as the Einstein universe, Minkowski 
spacetime, and 
the static open universe, respectively. However, it should be clear as well 
that by 
slicing, say, 
the $k=\pm 1$ solutions in an adequate way, the spherical and hyperbolic black 
holes can be described as states of the field theory on Minkowski space. This 
can be achieved more simply by choosing adequately the conformal factor 
between 
$h_{\mu\nu}$ and $\gamma_{\mu\nu}$, see \cite{horitz} for an example. We 
will make use of this idea later in section \ref{sec:rindler}. 

The case of $k=0$ is
particularly simple since, in the absence of any scale other than the thermal 
wavelength, conformal
invariance, together with staticity and homogeneity of the space, determines 
the stress tensor of the CFT to take the form
\beq
\langle {T_\mu}^\nu \rangle_{CFT}= {\sigma_{sb}\over \beta^{n}}{\rm 
diag}\left(-1,{1\over 
n-1},\dots,{1\over n-1}\right)\ .
\labell{strcft}
\eeq
The energy and entropy follow as
\beq 
E_{CFT}= \sigma_{sb} V_{n-1}\beta^{-n},\qquad
S_{CFT}={n\over n-1}\sigma_{sb} V_{n-1}\beta^{-n+1} \ .
\labell{escft}
\eeq
The factor
$\sigma_{sb}$ is the Stefan-Boltzmann constant, which is determined by
the precise field content of the CFT, and grows with the
number of degrees of freedom of the theory. We will give it below for 
the cases of interest. 

As observed in \cite{klts,wit1}, for planar ($k=0$) black holes 
$r_+\sim\beta^{-1}$, so the CFT thermodynamic functions \reef{strcft}, 
\reef{escft}, agree with their AdS black hole counterparts \reef{stressbh}, 
\reef{energy}, \reef{entropy} up to the Stefan-Boltzmann factors (notice that 
for $k=0$, $\epsilon^0_k=0$).  If one
wants to make this equivalence more precise and try to compare the precise 
Stefan-Boltzmann factors, then a specific dual field
theory has to be supplied.  String/M-theory provides duals for
AdS$_{n+1}$, $n=2,3,4,6,$ as the CFTs describing the world--volume 
dynamics of stacks of parallel (D1$+$D5)-, M2-, D3-, M5-branes. The 
dictionary for translating AdS/CFT quantities reads
\beqa
c={3l\over2G} \quad {\rm for\; AdS}_3\ ,&&\quad
N^{3/2} = {3\over 2\sqrt{2}} {l^2\over G} \quad {\rm for\; AdS}_4\ ,
\nonumber\\
N^2={\pi l^3\over 2 G}\quad {\rm for\; AdS}_5\ ,&&\quad
N^3 ={3\pi^2\over 16} {l^5\over G} \quad {\rm for\; AdS}_7\ ,
\labell{dictio}
\eeqa
where $N$ is the number of parallel branes. The powers of $N$ displayed 
above are measures of
the number of ``unconfined'' degrees of freedom:  for AdS$_5$, $N$ is the
rank of the gauge group of the dual ${\cal N}{=}4$ supersymmetric four
dimensional $SU(N)$ Yang--Mills theory. For AdS$_3$, $c$ is the central 
charge of the dual CFT in two dimensions; however, since
there are no $k=-1$ black holes in AdS$_3$ we will not deal with this case 
any longer.
Note that for generic number of dimensions, the entry in the dictionary can be 
expected to be
\beq
N^{n/2}\approx {l^{n-1}\over G}\quad {\rm for\; AdS}_{n+1}\ .
\labell{freedom}
\eeq

Let us focus now on the pair AdS$_5$/(${\cal N}{=}4$ SYM), in a discussion 
which can be traced back to \cite{guklpe}. Using 
\reef{dictio}, the results from \reef{energy} and \reef{entropy} for $k=0$, 
$n=4$, become
\beq
E_{BH}={3\pi^2 N^2\over 8 \beta^4}V_3 \ ,\quad S_{BH}={\pi^2 N^2\over 2 
\beta^3}V_3\ .
\labell{esbh}
\eeq
On the other hand, it is a standard result from {\it free} field theory at 
finite temperature that the 
factor $\sigma_{sb}$ in four dimensional thermal Minkowski space for fields of 
different spin is
\beq
\sigma_{sb}={\pi^2\over 30}\left(n_0+{7\over 4}n_{1/2} + 2n_1\right)\ ,
\labell{sbmink}\eeq
where $n_0$ is the number of (real) scalars, $n_{1/2}$ is the number of 
Weyl (or Majorana) fermions, and $n_1$ the number of gauge vectors. For 
${\cal N}{=}4$ $SU(N)$ SYM at large $N$,
\beq
n_0=6N^2\ ,\qquad n_{1/2}=4N^2\ ,\qquad n_0=N^2\ .
\labell{symcontent}
\eeq
By plugging these values into \reef{sbmink} we find $\sigma_{sb}=\pi^2 N^2/2$, 
which leads to the well--known result 
\cite{guklpe} that
\beq
E_{BH}={3\over 4} E_{SYM}\ ,\quad S_{BH}={3\over 4} S_{SYM}\ .
\labell{mismatch}\eeq
The SYM result is obtained by computing one--loop vacuum diagrams, \ie it 
is the leading term in a perturbative expansion in the 't~Hooft parameter 
$g_{YM}^2 N$. By contrast, the supergravity approximation, on which the AdS 
black hole result is based, is reliable only for large 
$g_{YM}^2 N$. The mismatch in \reef{mismatch} is therefore interpreted as 
a strong coupling effect. An argument for why the entropy should change only 
by a numerical factor of order one has been given in \cite{itzhaki}.

Let us comment on two aspects of \reef{mismatch}.  The first one is that the
values for the energy and entropy at strong coupling are smaller than their
perturbative values.  As a matter of fact, as noted in \cite{guklpe}, the 
result
for $E_{BH}$ would agree with a perturbative calculation if, for some reason, 
at
strong coupling we had effectively $n_0=6N^2$, $n_{1/2}=3N^2$, $n_0=0$, \ie if
only the scalar multiplets contributed to the free energy, whereas the fields 
in
the (${\cal N}{=}1$) vector multiplet could not be excited.  There is therefore
a {\it reduction} in the effective number of degrees of freedom at strong
coupling.  The second aspect we want to emphasize, for reasons which will be
better appreciated later, is that both the energy and the entropy are reduced 
by
the same factor $3/4$.  That this should happen is a consequence of the fact
that the temperature dependence $E\sim \beta^{-4}$ is the same at both strong
and weak coupling, since it is fixed by conformal invariance.  Therefore, even
if some degrees of freedom may get frozen at strong coupling, it appears that
all the states that contribute to the entropy also make a contribution to the
energy of the system.

For the cases of AdS$_4$ and AdS$_7$, the dual conformal field theories of $N$
parallel M2- and M5-branes are poorly known, and as a consequence it is
impossible at present to discuss these cases in the same detail as the
AdS$_5$/SYM pair.

Overall, we can say that the qualitative aspects of the AdS/CFT duality for
planar black holes are fairly well understood, and in particular for AdS$_5$ 
the
free field theory seems to capture a good deal of the thermodynamics at strong
coupling.

Turn now to $k=1$, \ie spherical black holes and their dual CFTs, which
according to \reef{bdry} are naturally defined on spatial spheres $S^{n-1}$.
This introduces a length scale $l$ in the theory.  As it happens, in this
instance there appears a phenomenon that is absent from planar (and hyperbolic)
AdS black holes.  The thermodynamic analysis of the black hole solutions 
reveals
a phase transition at finite temperature between the state corresponding to
$\mu=0$ (global AdS) and the (large) black hole phase \cite{hp,wit1,wit2}.  The
low temperature phase (global AdS) is interpreted as a ``confined" phase
\cite{wit1,wit2}.  This phenomenon, although expected from generic
considerations, can not be seen from a perturbative analysis of the field
theory.  Therefore, even if results for conformal fields on $S^1\times S^{n-1}$
at a perturbative level (free field theory) are available \cite{dowker}, which
can be employed to compute $E_{CFT}(\beta)$, they can not be expected to 
provide
us with any information about the strongly coupled regime, at least at low
temperatures:  the phase transition throws us into a region where perturbative
field theory is useless.

Nevertheless, there is one result that can be meaningfully compared, namely, 
the
Casimir energy associated to the field theory on $\IR\times S^3$.  The dual
supergravity solution is AdS$_5$ in global coordinates, which is protected from
strong coupling (string $\alpha'$) corrections \cite{kallosh}.  Moreover, the
Casimir energy is essentially determined by the central charges of the ${\cal
N}=4$ SYM theory, which receive no higher loop corrections \cite{anselmi}.
Indeed, it has been proven that the result from free field theory matches
precisely the AdS calculation \cite{pv}.

\section{AdS/CFT duality for hyperbolic black holes}\labels{sec:hyperduals}

Hyperbolic black holes share with planar black holes the property that they do
not exhibit phase transitions at finite temperature.  At any temperature the
phase structure is dominated by a black hole.  Then, the dual field theory at
strong coupling is expected to remain in an unconfined phase
\cite{re,birm}\footnote{It might be worth noting the following difference:  
Both
for the planar and the hyperbolic systems the free energy at any non--zero
temperature goes like $F\sim N^2$ (for AdS$_5$/SYM).  For planar black holes,
$\lim_{\beta\rightarrow\infty} \beta F=0$, and one could say that the phase
transition takes place at zero temperature, where the supergravity state is
AdS$_5$.  In contrast, in the hyperbolic case the phase at $T=0$ is still a
black hole (the extremal one), and, as we will see below,
$\lim_{\beta\rightarrow\infty} \beta F\sim N^2$.}.  Therefore, even if
interactions are expected to introduce modifications (as was the factor $3/4$ 
in
\reef{mismatch} for planar black holes), we can hope to be able to extract
valuable information by trying to connect the weakly coupled and strongly 
coupled regimes.

An important feature of hyperbolic black holes is that the curvature of the
hyperbolic space $H^{n-1}$ introduces a new scale into the field theory, and as
a result the temperature dependence is not fully fixed by conformal invariance.
The case of flat space is contained in this class of black holes as a limit (as
was also for spherical black holes), which can be characterized as the high
temperature limit.  At any other temperatures the thermodynamic functions are
more complicated, and encode more information than in the case of flat
spacetime.  In particular, the relationship between energy and entropy is not 
as
simple as in \reef{escft}, and the thermodynamic magnitudes become more
sensitive to the field theory content.

Furthermore, we will be able to crucially exploit a novel feature, absent from
the other two classes of black holes.  As mentioned above, there is one
particular state, the one corresponding to $\mu=0$ (\ie $r_+=l$, $\beta=2\pi
l$), which is isometric to AdS.  Since AdS$_5$($\times S^5$) is an exact string
state, protected from corrections in the `tHooft coupling $g^2_{YM}N$, results
at perturbative level can be extrapolated to strong coupling.  When we write 
AdS
with the hyperbolic slicing the situation is, however, interestingly
non--trivial, since the $k=-1$ description of AdS does not cover all of the
spacetime, rather only a wedge.  Accordingly, in a computation of, say, the
partition function of the theory, states that lie outside this wedge are traced
out, and will give rise to an entropy, sometimes called ``entanglement entropy"
\cite{entang}.  More precisely, on a Cauchy surface in the $k=-1$ patch, such 
as
shown Fig.~\ref{fig:patches}b, the data to the left of the Einstein-Rosen
bridge are traced out.  On the supergravity side, this entropy appears as an
entropy associated to the acceleration horizon.  On the field theory side we 
can
compute the entropy of states on a hyperbolic space.  The detailed comparison 
of
these quantities will only be possible for AdS$_5$/SYM, so we will devote most
of the section to this case.  Other sides of the relation to acceleration
horizons will appear in section \ref{sec:rindler}.

\subsection{AdS$_5$/SYM on a hyperboloid}

Let us then start by translating the strong coupling, black hole results of 
section \ref{sec:topobhs}
into field theory language by using \reef{dictio}, focusing on the AdS$_5$/SYM 
dual pair. From eqs.~\reef{stressads5}, \reef{rplus} we find, for the 
stress-energy tensor of strongly coupled SYM on hyperbolic space at finite 
temperature
\beq
\langle {T_\mu}^\nu \rangle_{\rm (sugra)}={\pi^2 N^2\over 
32\beta^4}\left(1+\sqrt{1+{2\beta^2\over \pi^2 l^2}}\right)^2  {\rm 
diag}(-3,1,1,1)\ .
\labell{stressstr}
\eeq
In this geometry, the energy is equal to $E=-\int d^3x \langle {T_0}^0 
\rangle$. 
On the other hand, from \reef{entropy}, the entropy is
\beq
S_{\rm (sugra)}={\pi^2 N^2 V_3\over 
16\beta^3}\left(1+\sqrt{1+{2\beta^2\over \pi^2 l^2}}\right)^3\ .
\labell{entrstr}
\eeq

It is straightforward to see that in the high temperature limit
$\beta\rightarrow 0$ we recover the results for flat space \reef{esbh}.

As explained, there are two states of particular interest. One is the 
extremal, 
zero temperature ($\beta\rightarrow\infty$) black hole \reef{ext4}, and the 
other is 
the solution isometric to AdS ($\beta=2\pi l$). For the first one we find 
\beq
E_{\rm (sugra)}|_{\beta\rightarrow\infty}=0
\labell{energround}
\eeq
and
\beq
S_{\rm (sugra)}|_{\beta\rightarrow\infty}={N^2\over 2^{5/2}\pi l^3}V_3\ .
\labell{entrground}
\eeq
Notice that the energy for this state, and actually the entire stress tensor,
is zero, so it seems appropriate to identify it with the ground state of the 
theory. Nevertheless, its entropy does not vanish, a surprising fact that was 
noted in \cite{ejm} and which we will discuss below.

For AdS$_5$ in the hyperbolic slicing, \ie the state at $\beta=2\pi l$,
\beq
E_{\rm (sugra)}|_{\beta=2\pi l}={3 N^2\over 32\pi^2 l^4}V_3\ ,
\labell{enerads}
\eeq
and
\beq
S_{\rm (sugra)}|_{\beta=2\pi l}={N^2\over 2\pi l^3}V_3 \ .
\labell{entrads}
\eeq

Turning now to the weakly coupled regime, we will make use of results obtained 
in \cite{candowk} for the stress tensor of conformal fields in $S^1\times 
H^3$. The essential 
input in the computation is the density of eigenvalues of the wave 
operator in $H^3$ for fields of different spins. If $h(s)$ is the number 
of helicities of the spin $s$ field, and $n_s$ is the number of such 
fields, then, for $s=0,1/2,1$, one gets
\beq
\langle {T_\mu}^\nu \rangle_{\rm (gauge)}=\sum_s {n_s h(s)\over 6\pi^2 
l^4}\int_0^\infty
d\lambda {\lambda (\lambda^2+s^2)\over e^{\beta\lambda/l}-(-1)^{2s}}\; {\rm 
diag}(-3,1,1,1)\ .
\labell{stressweak}
\eeq
The integrals can be performed explicitly\footnote{The reader should be 
aware that the integrations given in \cite{candowk} are not correct.}, and 
with $h(0)=1$, $h(1/2)=h(1)=2$ we find
\beq
\langle {T_\mu}^\nu \rangle_{\rm (gauge)}={\pi^2 \over 90\beta^4}\left( 
n_0+{7\over 
4}n_{1/2}+2n_1 +{5\beta^2\over 8\pi^2 l^2}(n_{1/2}+8 n_1)\right){\rm 
diag}(-3,1,1,1)\ .
\labell{stressweak2}
\eeq
Having the energy $E(\beta)$, the entropy can be 
computed by using the first law of thermodynamics, with the result
\beq
S_{\rm (gauge)}={2\pi^2 V_3\over 45\beta^3}\left( n_0+{7\over 
4}n_{1/2}+2n_1 +{15\beta^2\over 16\pi^2 l^2}(n_{1/2}+8 n_1)\right)\ .
\labell{entweak2}
\eeq
In the high temperature limit $\beta\rightarrow 0$ the results from the 
previous section for flat spacetime are recovered. However, 
attention should be drawn to the mixing of temperature dependences in 
\reef{stressweak2} and \reef{entweak2}. In 
contrast to the simple flat space dependence \reef{escft}, which would still 
hold if only scalar fields were present, the presence of higher spin fields 
introduces a sensitivity to the curvature of the space. This is reflected in 
the $\beta^2$ term inside brackets, which has a different factor in 
\reef{stressweak2} and \reef{entweak2}.

We specialize now to the field content of large $N$ $SU(N)$ ${\cal N}{=}4$ 
Super 
Yang--Mills theory, \reef{symcontent}, to find
\beq
\langle {T_\mu}^\nu \rangle_{\rm (gauge)}={\pi^2 N^2\over 6\beta^4}\left( 1 
+{\beta^2\over 
2\pi^2 l^2}\right){\rm diag}(-3,1,1,1)\ ,
\labell{stressweaksym}
\eeq
\beq
S_{\rm (gauge)}={2\pi^2 N^2\over 3\beta^3}V_3\left( 1 +{3\beta^2\over 4\pi^2 
l^2}\right)\ .
\labell{entweaksym}
\eeq
Compare now these results with the ones in the strongly coupled regime, eqs.\ 
\reef{stressstr}, \reef{entrstr}. It is apparent that the dependence on the 
temperature is rather different, and in fact both come to agree 
only at high temperatures, where we recover the same relationship as in 
\reef{mismatch}.

For the ground state at zero temperature, the results
\beq
E_{\rm (gauge)}|_{\beta\rightarrow\infty}=S_{\rm 
(gauge)}|_{\beta\rightarrow\infty}=0
\labell{enerweak0}\eeq
are as expected for a conventional ground state. On the other hand, for 
the state at $\beta=2\pi l$ we get
\beq
E_{\rm (gauge)}|_{\beta=2\pi l}={3 N^2\over 32\pi^2 l^4}V_3 \ ,
\labell{enerweakads}
\eeq
\beq
S_{\rm (gauge)}|_{\beta=2\pi l}={N^2\over 3\pi l^3}V_3\ .
\eeq

It is immediate to notice that for both the ground state and the state at
$\beta=2\pi l$ the energy computed using free field theory agrees with the
results in the strong coupling (supergravity) regime, eqs.~\reef{enerads},
\reef{energround}.

As a matter of fact, not only in supergravity but also in the field theory on
$S^1\times H^{n-1}$ the state at $\beta=2\pi l$ is singled out among states at
other temperatures:  it can be formally obtained from the vacuum of the 
Einstein
universe $\IR\times S^{n-1}$ by ``thermalization at imaginary temperature"
$T=(2\pi il)^{-1}$ \cite{candowk}.  The two calculations of (a) the Casimir
energy on $\IR\times S^3$ \cite{pv}, and (b) the energy of the state at
$\beta=2\pi l$ on $S^1\times H^3$, eqs.~\reef{enerweakads} and \reef{enerads},
are in this light seen as the result of formally equivalent calculations.  This
is reflected in the fact that both follow from the same central charge of the
field theory.

Despite the agreement for the energies of these states, the results for the 
entropy obtained from supergravity are both different from the one--loop 
field theory results. Eq.~\reef{entrground} is telling us that at infinite 
`tHooft 
coupling there is a large degeneracy for the state at zero temperature and zero 
energy density. Such ground state degeneracies are highly unusual. 
The mismatch in the entropy for the state at $\beta=2\pi l$ is not less 
unexpected. As we had remarked, this state is described in the bulk of AdS as a 
wedge of the full AdS$_5$ spacetime. We shall argue in sec.~\ref{sec:correc} 
that not only the energy, but also the entropy of this state would have been 
expected to be protected from corrections in the coupling. Nevertheless, we 
find 
at strong coupling an entropy {\it larger} than that obtained from field theory 
at the lowest perturbative order.  The relationship between both is
simple,
\beq
S_{\rm (sugra)}|_{\beta=2\pi l}={3\over 2} S_{\rm (gauge)}|_{\beta=2\pi l}\ .
\labell{smismatch}
\eeq

This is in stark contrast to the situation for planar black holes, where there 
is
an effective reduction at strong coupling in the number of states available to
the gauge theory.  Here we find instead an enhancement, but one that affects 
only
the entropy, not the energy\footnote{It is amusing to observe, although we do 
not
mean to attach too much significance to this remark, that the mismatch between
entropies at $\beta=2\pi l$, eq.~\reef{smismatch}, could be remedied by 
assuming
that, at that particular temperature, there were $4N^2$ additional chiral
multiplets, $\delta n_0=8N^2$, $\delta n_{1/2}=4N^2$, contributing to the 
entropy
but not to the stress tensor.}.

It can be readily checked that the values of the energy at weak and strong 
coupling agree only for the two values of the temperature $\beta=2\pi l,\; 
\infty$. Indeed, we would not have expected agreement at any other 
temperature, due to strong coupling corrections. It is therefore difficult 
to meaningfully make comparisons of the entropy expected at different 
temperatures. Notice, however, that, at any temperature, 
\beq
{S_{\rm (gauge)}(\beta)\over E_{\rm (gauge)}(\beta)}\leq {S_{\rm 
(sugra)}(\beta)\over E_{\rm 
(sugra)}(\beta)}\ ,
\eeq
with equality only at infinite temperature. Although the different temperature 
dependence of different fields makes it difficult to take this too literally, 
this inequality would suggest that at strong coupling there appear to be more 
states contributing to the entropy than those that contribute to the energy.

One last magnitude which is interesting for comparison purposes is the 
specific heat, since it measures the response (susceptibility) of the degrees 
of freedom to thermal excitation. We find 
\beq
C_{\rm (sugra)}={3\pi^2 N^2\over 16 \beta^3}{\left(1+\sqrt{1+{2\beta^2\over 
\pi^2 l^2}}\right)^3 \over \sqrt{1+{2\beta^2\over \pi^2 l^2}}}\ ,
\labell{cstrong}
\eeq
\beqa
C_{\rm (gauge)}&=&{2\pi^2\over 15\beta^3}\left( n_0+{7\over 
4}n_{1/2}+2n_1 +{5\beta^2\over 16\pi^2 l^2}(n_{1/2}+8 n_1)\right)
\nonumber\\
&=&{2\pi^2 N^2\over \beta^3}\left(1 +{\beta^2\over 4\pi^2 l^2}\right)\ .
\labell{cweak}
\eeqa

For the different temperatures of interest these become
\beq
C_{\rm (gauge)}|_{\beta\rightarrow\infty}\rightarrow {4\over 3} C_{\rm 
(sugra)}|_{\beta\rightarrow\infty}\rightarrow {N^2\over 2 l^2 \beta}\ ,
\labell{clotemp}
\eeq
\beq
C_{\rm (gauge)}|_{\beta=2\pi l}=C_{\rm (sugra)}|_{\beta=2\pi l}={N^2\over 2\pi 
l^3}\ ,
\labell{cloads}
\eeq
while at high temperature we recover the flat space result
\beq
C_{\rm (gauge)}|_{\beta\rightarrow 0}\rightarrow {4\over 3} C_{\rm 
(sugra)}|_{\beta\rightarrow 0}\rightarrow {2 \pi^2 N^2\over \beta^3}\ ,
\labell{chitemp}
\eeq
in which the specific heat grows with the characteristic four-dimensional 
dependence $\sim\beta^{-3}=T^3$. 

At low temperatures it is the spin-1/2 and spin-1 fields which dominate the
specific heat, at least at weak coupling (see \reef{cweak}).  The curvature of
$H^3$, to which these fields are sensitive, makes them more susceptible of 
being
excited and as a consequence the specific heat grows faster, as $C\sim T$
instead of $T^3$.  Remarkably, this is also the behavior we find at strong
coupling, where we do not know how to separate the contributions from different
sets of degrees of freedom.  This suggests that, at low temperatures, the
degrees of freedom at strong coupling are not too dissimilar in nature from
those that operate at weak coupling.  Curiously, the precise numerical factor 
is
off by the same fraction ${4/3}$ as at high temperatures.

Finally, at $\beta=2\pi l$ the specific heats are exactly the same in the
strongly coupled and weakly coupled regimes.  It would appear that even if 
extra
states are present which contribute to the entropy (albeit not to the energy),
the susceptibility to thermal excitation is still dominated by those states 
that
make up the energy density.

\subsection{Other dimensions}

The results \reef{energy} and \reef{entropy} for the energy and entropy of 
topological black holes in section 
\ref{sec:topobhs} can be expressed in terms 
of the temperature using \reef{rplus} and then converted into expressions for 
field theory at strong coupling using the dictionary \reef{dictio} or, more 
generally, 
\reef{freedom}. In turn, it is possible as well to compute the corresponding 
quantities for free fields on hyperbolic space at finite temperature. The 
contribution from a spin $s$ field to the energy on $S^1\times H^{n-1}$ is 
obtained as
\beq
E_s(\beta)=h(s){V_{n-1}\over \omega_{n-1}l^{n}}\int_0^\infty
d\lambda {\lambda \mu_s(\lambda)\over e^{\beta\lambda/l}\mp 1}\ ,
\labell{ens}
\eeq
where $h(s)$ is, as before, the number of physical states (helicities) for 
each 
field, $\omega_{n-1}$ is the volume of the unit $(n-1)$-sphere, the $-$ ($+$) 
sign in the denominator applies to boson (fermion) fields, and $\mu_s(\lambda)$ 
measures the 
degeneracy (density) of eigenvalues of the wave operator on the hyperbolic 
space. It is known in the mathematical literature as the Plancherel measure. 
The 
latter has been computed for hyperbolic spaces in arbitrary dimensions for a 
wide variety of fields. To quote, for $H^{N}$, for real scalars 
\cite{campo},
\beq
\mu_{sc}(\lambda)={\pi \over [2^{N-2}\Gamma(N/2)]^2} \left|{\Gamma(i\lambda 
+{N-1\over 2})\over \Gamma(i\lambda)}\right|^2\ .
\eeq
For spinors \cite{camposp},
\beq
\mu_{sp}(\lambda)={\cosh\pi\lambda \over [2^{N-2}\Gamma(N/2)]^2} 
\left|\Gamma\left(i\lambda +{N\over 2}\right)\right|^2\ .
\eeq
For (co-exact) $p$-forms, $h(s)={(N-1)!\over p!(N-p-1)!}$ (which must be 
halved 
for self-dual forms), and \cite{campopform}
\beq
\mu_{p-{\rm form}}(\lambda)={\pi \over [2^{N-2}\Gamma(N/2)]^2}{1\over 
\lambda^2 
+ ({N-1\over 2}-p)^2} \left|{\Gamma(i\lambda +{N+1\over 2})\over 
\Gamma(i\lambda)}\right|^2\ ,
\eeq
(this contains the scalar case for the value $p=0$, and gauge vectors for 
$p=1$).

Using these results\footnote{These results 
have been used in \cite{byts} to perform calculations similar to those 
described here.}, we can compute for the free field content of a {\it 
single} M2-brane, \ie an ${\cal N}=8$ supermultiplet in $d=3$,
\beq
E_{M2}(\beta)={V_2\over 4 l^3}\int_0^\infty
d\lambda\; \lambda^2\left({n_0 \tanh\pi\lambda\over e^{\beta\lambda/l}- 1}+
{n_{1/2} \coth\pi\lambda\over e^{\beta\lambda/l}+1}\right) \ ,
\labell{m2}
\eeq
with $n_0=8$ and $n_{1/2}=8$. It is not possible to give the results of the 
integrations in closed form for arbitrary values of $\beta$, although they 
simplify for $\beta=2\pi l$. However, it is easy to see that these results 
bear 
little resemblance to the ones that follow from supergravity 
calculations. Indeed, from \reef{m2} one easily sees that
\beq
E_{M2}|_{\beta\rightarrow\infty}=S_{M2}|_{\beta\rightarrow\infty}=0\ ,
\eeq
whereas the strong coupling calculation would yield
\beq
E_{M2(\rm strong)}|_{\beta\rightarrow\infty}=-{N^{3/2}V_2\over 9\sqrt{6} \pi 
l^3}\ ,\quad S_{M2(\rm 
strong)}|_{\beta\rightarrow\infty}= {N^{3/2}V_2\over 9\sqrt{2} l^2}\ .
\eeq
Notice that not only the entropy but also the energy of this 
state is different from zero. Indeed, as noted in 
\cite{ejm}, for AdS$_4$ (and in fact for all 
even $d$ AdS$_{d}$) it is the state at $\beta=2\pi l$ that has zero energy. 
And nevertheless it has non--zero entropy. 

These results for AdS$_4$ are even more striking 
than those for AdS$_5$: the state that is isometric to AdS$_4$ is a state at 
finite temperature and with non--zero entropy, which nevertheless 
has zero energy density! This looks markedly 
different from conventional field theory. Moreover, there appear states, 
namely, those for $\mu_e\leq \mu <0$, with 
{\it total} negative energy. The meaning of these is unclear.

For the free field content of the $(2,0)$ superconformal theory on a single 
M5-brane (\ie the 
$d=6$ tensor supermultiplet), 
\beqa
E_{M5}(\beta)&=&{V_5\over 36\pi^3 l^6}\int_0^\infty
d\lambda\; \lambda\left({5 \lambda^2(\lambda^2+1)\over e^{\beta\lambda/l}- 1}+
{8 (\lambda^2+1/4)(\lambda^2+9/4)\over e^{\beta\lambda/l}+1}+{3 
(\lambda^2+1)(\lambda^2+2)\over e^{\beta\lambda/l}- 1}\right)\nonumber\\
&=&{V_5 \pi^3 \over 1440\beta^6}\left(80+84{\beta^2\over \pi^2 
l^2}+55{\beta^4\over \pi^4 l^4}\right)\ .
\labell{m5}
\eeqa
This free field theory expression has nothing exotic about it. As the 
temperature goes to zero, the energy vanishes. So does the entropy, too, as 
can be checked easily. However, the 
results at strong coupling from AdS calculations are puzzling: neither the 
state at zero temperature nor the one at $\beta=2\pi l$ have zero energy. 
Both, moreover, have non--zero entropy. 

It appears quite likely that in all these cases exotic states that contribute 
to the entropy but not to the energy would be required to account for the black 
hole entropy. However, and in contrast to the case of AdS$_5$/SYM, for AdS$_4$ 
and AdS$_7$ free field theory yields little useful information.

\section{Dual CFT description in Rindler space}\labels{sec:rindler}

In the previous section we have chosen to describe hyperbolic black holes in 
terms of 
the theory on $\IR\times H^{n-1}$. However, as noted at the beginning of  
section \ref{sec:review}, it is possible to perform the entire description 
in 
terms of the theory on flat space. All one has to do is slice the AdS black 
hole 
spacetime in a way that, near the boundary, the constant radius sections are 
flat. Let us start by showing 
how this can be done explicitly for the solutions which are locally isometric 
to AdS$_{n+1}$. In Poincar\'e coordinates,
\beq
ds^2={r^2\over l^2}(-du\; dv +dx_i^2)+{l^2\over r^2}dr^2
\eeq
($i=1,\dots,n-2$), where $u,v$ are light-cone coordinates $u=t-x_{n-1}$, 
$v=t+x_{n-1}$. Now change
\beq
u=-\zeta\sqrt{1-{l^2\over{\tilde r}^2}}\; e^{-\eta/l}\ ,\quad 
v=\zeta\sqrt{1-{l^2\over{\tilde r}^2}}\; e^{\eta/l} \ ,
\labell{changeuv}
\eeq
\beq
r=l{{\tilde r} \over \zeta}\ ,
\labell{changer}
\eeq
leaving $x_i$ unchanged, to find AdS in hyperbolic ($k=-1$) coordinates
\beq
ds^2=-\left({{\tilde r}^2\over l^2}-1\right)d\eta^2 +{d{\tilde r}^2\over 
{{\tilde 
r}^2\over 
l^2}-1}+{\tilde r}^2 {d\zeta^2 +dx_i^2\over \zeta^2}
\eeq
(the spatial hyperboloid $H^{n-1}$ is parametrized here in horospheric 
coordinates). The Killing horizon at ${\tilde r}= l$ is mapped onto the null 
surfaces 
$u,v=0$. Now notice that as the boundary is approached 
($r,{\tilde r}\rightarrow\infty$) the transformation \reef{changeuv} between 
boundary coordinates becomes
\beq
u\rightarrow-\zeta e^{-\eta/l}\ ,\quad v\rightarrow\zeta e^{\eta/l} \ .
\eeq
This is precisely the transformation between Minkowski and Rindler 
coordinates. Indeed, as we approach the boundary,
\beqa
{r^2\over l^2}(-du\; dv +dx_i^2)&\rightarrow &{{\tilde r}^2\over 
l^2}\left(-d\eta^2+l^2{d\zeta^2 +dx_i^2\over \zeta^2}\right)\nonumber\\
&=&{r^2\over l^2}\left(-\zeta^2 \left({d\eta\over l}\right)^2+d\zeta^2+dx_i^2 
\right)={r^2\over l^2}ds^2_{R}\ ,
\labell{hypertorind}
\eeqa
where $ds^2_{R}$ is the metric on Rindler space (with time 
rescaled by $l$). The conformal factor that effects the change between  
$\IR\times H^{n-1}$ and the flat (Rindler) space at the asymptotic boundary is 
$r^2/{\tilde r}^2=l^2/\zeta^2$.

Therefore AdS in hyperbolic coordinates corresponds, in the description in
terms of a field theory in flat space, to the Rindler state of the CFT at
$\beta=2\pi l$.  This is, of course, the Rindler description of the Minkowski
vacuum. For the Rindler observer, the latter is a mixed thermal state described 
by a density matrix. 

This relationship between the solutions that are isometric to AdS, and
their corresponding states in the dual field theories, is entirely analogous to
that existing between BTZ black holes and AdS$_3$ and the corresponding states
in the dual $1+1$ CFTs \cite{anju}, except for the fact that we have not 
performed discrete identifications.

For black holes with $\mu\neq 0$ it is not simple to find a global coordinate
transformation that effects the change from the hyperbolic to the flat space
description.  However, we only need the conformal factor that transforms the
boundary geometries near infinity, and then use it to transform the stress
tensors as in eq.~\reef{tensors}.  This was the procedure followed in
\cite{horitz} to find a description of spherical black holes in terms of the 
SYM
theory in Minkowski space.  We can do the same thing here using the conformal
factor $l^2/\zeta^2$, which according to eq.~\reef{hypertorind} takes us to a 
Rindler space geometry at the boundary.

We conclude then that, in the dual field theory on flat space, hyperbolic black
holes correspond to thermal Rindler states at temperature $\beta$.  The one at
temperature $\beta=2\pi l$ is singled out as corresponding to the Poincar\'e
invariant vacuum in Minkowski space.  Other states are described, in
imaginary time, using geometries with a conical singularity at $\zeta=0$.
Notice, however, that there is no conical singularity in the description on
$\IR\times H^{n-1}$.

Given the conformal factor between hyperbolic space and Rindler space, the 
stress tensor in the latter is constructed as
\beq
{\langle{T_\mu}^{\nu}\rangle}^{\rm (Rindler)}(\beta)={l^4\over 
\zeta^4}\left[ {\langle{T_\mu}^{\nu}\rangle}^{\rm (Hyper)}(\beta) - 
{\langle{T_\mu}^{\nu}\rangle}^{\rm (Hyper)}(\beta=2\pi l)\right]
\labell{rindhyp}
\eeq
(we are implicitly using the fact that the trace anomaly vanishes for 
both spaces). As a matter of fact, the simple conformal relationship between 
Rindler and hyperbolic space has been put to use before in order to solve one 
of 
them from knowledge of the other \cite{candowk} (see also, \eg 
\cite{merind,moretti} and references therein). Conventionally, we have 
subtracted a $\beta$-independent tensor term in \reef{rindhyp} in order that 
that the stress tensor vanishes at $\beta=2\pi l$.
The reason is that $\langle T_{\mu\nu}\rangle$ being a tensor that vanishes in 
the Minkowski vacuum, it should vanish in that state in any other coordinate 
system. Recall that the Minkowski vacuum is the global state of minimum energy, 
which realizes the full Poincar\'e symmetry of the theory. The Rindler vacuum 
(the state for $\beta\rightarrow\infty$) can have lower energy because the 
minimization of the energy in Rindler space is constrained only by a subgroup 
of the Poincar\'e symmetries. The divergence of the vacuum energy density as 
$\zeta\rightarrow 0$ is only expected, since the vacuum is made to accelerate 
infinitely hard at that point.

The construction \reef{rindhyp} works the same way at weak and strong coupling.
Since at $\beta=2\pi l$ the stress tensor ${\langle{T_\mu}^{\nu}\rangle}^{\rm
(Hyper)}$ is the same in both regimes, the negative energy of the Rindler vacuum
is the same at zero and infinite coupling.  Moreover, the subtraction of such a
quantity does not affect the calculation of the entropy.  Therefore, the
subtraction in \reef{rindhyp} does not introduce any significant modification in
our discussion.

The entropy density $s$ in Rindler space is equally obtained by rescaling the 
one in hyperbolic space as
\beq
s^{\rm (Rindler)}(\beta)={l^3\over \zeta^3}s^{\rm (Hyper)}(\beta)\ .
\eeq
Notice that the finite entropy density of the Minkowski vacuum is not to be 
subtracted, since it is physically relevant as entanglement entropy. A 
Minkowski 
(global) observer assigns zero entropy to this state.  An
accelerating observer, however, detects quantum fluctuations of the vacuum
(they appear to him as thermal fluctuations) and is sensitive to the vacuum
activity of the global fundamental state.  The Rindler entropy density at 
$\beta= 2\pi l$ therefore yields a measure of the states that are subject to 
quantum fluctuations in the global vacuum.

The discussion of the previous section can now be couched in terms of statements
about the energy and entropy of the CFT in Rindler space.  We therefore find
that the Rindler vacuum is, at infinite coupling, highly degenerate.  On the
other hand, a Rindler observer accelerating in the Minkowski vacuum measures an
entropy density for SYM larger than would have been expected from the weak
coupling calculation.

\section{Finite 't~Hooft coupling corrections}\labels{sec:correc}

The calculations we have presented so far have been performed at two opposite
ends of the scale of the SYM 't~Hooft coupling, $g^2_{YM} N$.  Gauge theory
computations have been performed at the level of one--loop vacuum diagrams, \ie
$g^2_{YM} N=0$, whereas the supergravity approximation to type IIB string 
theory
is reliable when $g^2_{YM} N=l^4/(2 \alpha'^2) \rightarrow \infty$.  In this
section we want to discuss the corrections that arise when $g^2_{YM} N$ 
deviates from these limits. The study will be carried out only for the case of 
AdS$_5$/SYM.

Conformal invariance imposes strong restrictions on the form of finite coupling 
corrections for the planar case, $k=0$. The temperature dependence is fixed, so 
a thermodynamic function like the free energy must be of the form
\beq
F=F_0\; f(g^2_{YM} N)\ ,
\eeq
where $F_0$ is the value at zero coupling. It is clear that the energy and 
entropy are corrected by the same function $f(g^2_{YM} N)$. In contrast, for 
the 
hyperbolic or spherical systems one will typically have
\beq
F=F_0\; f(g^2_{YM}N,\beta/l)\ ,
\eeq
and the temperature dependence will change in general. Indeed, we have seen 
explicitly that the supergravity and gauge theory expressions for the energy 
and entropy in the hyperbolic case have a very different dependence on $\beta$. 
One would ascribe the differences to the effect of interactions as the coupling 
is turned on.

Perturbative interactions will change the weak coupling result through 
higher--loop diagrams. For the SYM theory in Minkowski space, these have been 
computed in \cite{loops}. However, the extension of these calculations to the 
spherical or hyperbolic cases is much more difficult, since it implies 
solving an interacting theory in a curved background.

At the other end of the scale, large $g^2_{YM} N$, the first corrections arise 
from $O(\alpha'^3)=O((g^2_{YM}N)^{-3/2})$ corrections to the effective IIB 
superstring action at low 
energies. The relevant term in the Euclidean action is \cite{strcorr,planar1}
\beq
\delta I=-{1\over 16\pi G_{10}}\int d^{10}x \sqrt{g^{(10)}}
\alpha'^3{\zeta(3)\over 8} W\ ,
\labell{deltai}
\eeq
where $W$ is a scalar constructed out of contractions of four Weyl tensors. 
Using this term, finite coupling corrections to the thermodynamics of planar 
black holes have been studied in \cite{planar1,planar2}, and spherical black 
holes in \cite{spheric1,spheric2}. The study of the latter has been taken 
further in \cite{calklem}, where the corrections to hyperbolic black holes have 
been calculated as well.

In principle, one must consider corrections in the entire ten dimensional
theory, since it is not possible to keep the size of the sphere $S^5$ fixed
\cite{planar2}.  However, on reduction to five dimensions it is easy to see
that neither the dilaton nor the scale factor of the sphere will contribute
on--shell to the effective five--dimensional action for as long as they fall
off fast enough at asymptotic infinity.  It is therefore possible to compute
the Euclidean action of the corrected solutions in a five dimensional
formulation for solutions asymptotic to AdS$_5$ \cite{planar1}.  This will be
important for us, since it will permit us to employ the intrinsic
regularization procedure of \cite{pv} for the computation of the corrected
action.

Let us focus on the hyperbolic solution at $\beta=2 \pi l$.  The full
ten--dimensional solution is locally AdS$_5\times S^5$, which is conformally
flat.  Therefore corrections from \reef{deltai} vanish and the geometry should
remain the same.  Actually, it appears reasonable to assume that all $\alpha'$
corrections can be written in an appropriate scheme in terms of the Weyl tensor 
(along the lines in \cite{ambig}). It then follows that the temperature 
$\beta=2\pi l$ is 
uncorrected, since it is
determined entirely by the properties of the metric.  If we use the intrinsic 
regularization of the gravitational action we have employed in this paper, 
which relies only on the metric of the solution at hand, then the value of the 
action for this solution is also unchanged. Now, since the Euclidean action is 
identified with $\beta F$, it follows that the free energy of the state at 
$\beta=2\pi l$ should receive no corrections.  We have already mentioned that, 
from field theory arguments, the energy of this state is protected, and indeed 
we have explicitly seen that it takes the same value at zero and infinite 
coupling.  Now, when higher derivative terms are added to the Einstein--Hilbert 
action the entropy is in general no longer given by the area. However, since
\beq
S=\beta(E-F)\ ,
\eeq
we would conclude that the entropy of SYM on hyperbolic space at $\beta=2\pi l$ 
should not change its value when going from strong to weak 
coupling\footnote{Indeed, using an intrinsic regularization procedure it 
appears that any thermodynamic quantity that can be computed solely from 
properties of the metric of the solution considered should remain 
uncorrected.}. {\it This 
is not what we have found.} The entropy at strong coupling is instead $3/2$ 
times larger than the value computed from one--loop vacuum diagrams. Obviously, 
already the free energy is different in both regimes,
\beq
F_{\rm (sugra)}|_{\beta=2\pi l}= -{5 N^2\over 32 \pi^2 l^4}V_3=
{15\over 7} F_{\rm (gauge)}|_{\beta=2\pi l}\ .
\eeq

This looks worrisome.  While we cannot completely discard that subtle reasons
invalidate the assumption that the higher $\alpha'$ corrections can be written
in some scheme in terms of the Weyl tensor, it should be noted that the argument
developed above is known to actually {\it work} for the closely analogous
situation of BTZ$\times S^3$ black holes \cite{planar1}.  It would certainly
seem odd if the free energy were corrected in the $\alpha'$ expansion, but the
energy density (and specific heat) were not.  Let us then discuss other
alternatives here.  In concluding that the entropy should remain unchanged we
have implicitly assumed that there is no phase transition in the theory as a
function of the coupling.  It might then be that as the coupling is increased a
phase transition occurs, in which new states arise that do not change the energy
but nevertheless increase the entropy.  This phase transition would have to be
invisible in an expansion of $F$ in inverse powers of $g^2_{YM}N$.  Another
possibility, probably no less bizarre, is that the one--loop calculation at weak
coupling does not capture all of the states that build up the entropy.  If this
were the case, the correct result to all orders would be the one given by the
supergravity calculation.  We will discuss further this possibility later in
sec.~\ref{sec:precursor}.

States other than the one at $\beta=2\pi l$ are expected to receive 
corrections. 
These should change the entire ten--dimensional metric, and with it the 
temperature and thermodynamic functions. Indeed, these corrections have been 
computed in \cite{calklem}, where it has been calculated how the value of $r_+$ 
as a function of $\beta$ is shifted. Although the value $r_+=l$ for $\beta=2\pi 
l$ remains, as argued, uncorrected, the extremal radius changes. As for the 
action we get 
\beqa
\delta F&=&-{15 \pi^2\zeta(3)\over 128} {V_3 \alpha'^3 \over G l^{3}\beta^4} 
\left(1+k{\beta^2\over \pi^2 l^2}{1\over 1+\sqrt{1-k{2\beta^2\over \pi^2 
l^2}}}\right)^4
\nonumber \\
&=&-{15\pi^2 \zeta(3)\over 64} {N^2 V_3 \over \beta^4}(2g^2_{YM} N)^{-3/2} 
\left(1+k{\beta^2\over \pi^2 l^2}{1\over 1+\sqrt{1-k{2\beta^2\over \pi^2 
l^2}}}\right)^4
\labell{deltai2}
\eeqa
(here $G$ is the five--dimensional Newton's constant).  We have performed the
calculation of the corrected action using, as in the rest of this paper, the
intrinsic regularization method of \cite{pv}.  The calculation of the action in
\cite{calklem} was instead done with a background subtraction.  Our result
coincides with the one in \cite{calklem} except for one important difference:
for $k=-1$ the value of $\delta F$ in \reef{deltai2} does not tend to zero as
the temperature goes to zero, $\beta\rightarrow\infty$.  This is, the energy of
the extremal state is shifted from zero.  In contrast, the calculations in
\cite{calklem} were performed by taking the state at zero temperature as the
reference state.  By construction, this keeps the energy of that state to zero.
But this way of proceeding has the unattractive property that the energy of the
state at $\beta=2\pi l$ does receive a correction to this order.  It would seem
unnatural to choose a regularization that needlessly introduces finite coupling
corrections for a quantity that we have reasons to expect should remain 
uncorrected. Intrinsic regularization yields instead $\delta F=0$ at $\beta=2\pi 
l$.

It is interesting to observe that for $k=-1$ the corrections change sign at the 
AdS value $\beta=2\pi l$. Using \reef{deltai2} it is a straightforward matter 
to compute the corrections to the energy, entropy, and specific heat. The 
explicit formulae for arbitrary temperature are rather unilluminating, so we 
shall only quote the values for the states of most interest. Of course,
\beq
\delta E=\delta S= \delta C=0 \quad {\rm at}\,\, \beta =2\pi l\ ,
\eeq
while
\beqa
\delta E|_{\beta\rightarrow\infty}&=&-{15\zeta(3) \over 256 \pi^2 l^4} N^2 V_3 
(2g^2_{YM} N)^{-3/2}\ ,\nonumber\\
\delta S|_{\beta\rightarrow\infty}&=&-{45\zeta(3) \over 64 \sqrt{2} \pi l^3} 
N^2 V_3 
(2g^2_{YM} N)^{-3/2}\ ,\nonumber\\
\delta C|_{\beta\rightarrow\infty}&\rightarrow&{105\zeta(3) \over 32 l^2 \beta} 
N^2
(2g^2_{YM} N)^{-3/2}\ .
\eeqa
Notice that the extremal state acquires a negative energy. At present it does 
not seem possible to decide whether this is a real problem or just an artifact 
of the $\alpha'$ expansion. It can be made to appear less problematic by taking 
the Rindler interpretation of the result, since it merely implies a shift in the 
energy of the Rindler vacuum of SYM.
The correction to the entropy is negative as well. 
Extrapolation is not admissible at this level, but it might be that the large 
degeneracy of the ground state at infinite $g^2_{YM} N$ steadily decreases with 
the coupling. Perhaps more significant is the fact that the corrections to the 
specific heat maintain the dependence $C\sim \beta^{-1}$ that we have seen 
already appears for higher spin fields in hyperbolic space.

\section{Discussion}\labels{sec:precursor}

We hope to have made it clear that hyperbolic black holes provide a rich
setting to study the AdS/CFT correspondence, introducing new features absent
from both planar and spherical black holes.  A particularly interesting aspect
is that they provide the possibility of studying properties of the global AdS
vacuum and of the Minkowski vacuum of the CFT by the introduction of
accelerating observers.

The most striking result of our analysis has been the identification of
enhancements in the value of the entropy that are not accompanied by increments
in the energy.  The first instance of this phenomenon is the appearance of a
large degeneracy for the ground state at infinite coupling.  Large degeneracies
for supersymmetric, zero temperature black holes are well known in string
theory\footnote{They arise as degeneracies of BPS states.}.  However, the
hyperbolic extremal black hole is not supersymmetric, and in the absence of
supersymmetry it is extremely difficult to make interacting systems have highly
degenerate ground states\footnote{A somewhat similar phenomenon has been found
for charged AdS black holes \cite{cejm}.  However, in that case these large
degeneracies are accompanied by equally large energy densities, and the states
are moreover known to be unstable.}.  It may be worth recalling that the result
can be interpreted as saying that the Rindler vacuum of SYM at infinite 
't~Hooft
coupling is highly degenerate.

No less unexpected is the strong/weak coupling discrepancy of the entropy of 
AdS
in hyperbolic slicing.  This time it can be interpreted in terms of the
degeneracy of the Minkowski vacuum of SYM as seen by an accelerating observer.
The mismatch in the entropy is the more striking, since we did not expect
corrections to the free energy of this state at any order.  Indeed, we have
explicitly seen that there are no corrections to $O(\alpha'^3)$, and that the
energy and specific heat of that state take the same value at zero and infinite
coupling.  Barring subtleties in the $\alpha'$ expansion, alternative
explanations must be sought.  We have mentioned the possibility that the states
responsible for this entropy arise as a consequence of a phase transition as 
the
coupling is increased.  Another option might be that the non--renormalization 
of
the entropy still works, but that the total entropy at small coupling is not
entirely captured by standard one--loop vacuum diagrams, \ie that the
Super-Yang-Mills theory possesses states that contribute to the entropy but not
to the energy density.  This would sound like a rather exotic proposal.
However, very similar conclusions have been arrived at in \cite{pst}, from the
study of an entirely different paradox in the AdS/CFT context.  There, in order
to preserve causality of the field theory when describing processes that take
place far from the boundary of AdS, it was found necessary to postulate ``{\sl 
a
very rich collection of hidden degrees of freedom of the SYM theory which store
information but give rise to no local energy density}" ({\sl sic}) \cite{pst}.
It is striking that this appears to be the sort of phenomenon we are observing
in our study of black hole entropy.  From the arguments in \cite{pst} it would
appear that these so--called ``precursor" states are already present at the
weakly coupled level, and therefore might provide the extra degeneracies we 
have
found.

As noted, even if AdS$_4$ and AdS$_7$ also appear to exhibit enhanced 
entropies,
the situation is complicated by the lack of an adequate understanding of their
dual field theories.  It will be obviously interesting to find other setups
where these exotic entropies show up.

\section*{Acknowledgements}

We are grateful to Andrew Chamblin, Ivo Sachs, and Manuel Valle-Basagoiti for
interesting conversations.  We are particularly indebted to Clifford Johnson
and Rob Myers for many suggestions and detailed comments on an earlier version
of the manuscript.  Our thanks as well to Dietmar Klemm for bringing
ref.~\cite{calklem} to our attention.  Work supported by EPSRC through grant
GR/L38158 (UK), and by grant UPV 063.310-EB187/98 (Spain).

\end{document}